\documentstyle[eqsecnum,aps,prl]{revtex}
\tighten

\begin{document}

\draft
\twocolumn
\preprint{HU-TFT-95-20}

\title{Diffusion Processes and Growth on Stepped Metal Surfaces}

\author{J. Merikoski$^{1,}$\cite{PerAdd} and T. Ala--Nissila$^{1,2}$}
\address{$^1$Research Institute for Theoretical Physics,
P.O. Box 9, FIN--00014 University of Helsinki, Finland\\
$^2$Department of Physics, Brown University, Box 1843, Providence
R.I. 02912, U.S.A., and\\
Tampere University of Technology, P.O. Box 692,
FIN--33101 Tampere, Finland}

\date{15 May 1995 (to appear in Phys. Rev. B Rapid Comm.)}
\maketitle

%==================================================================== ABSTRACT

\begin{abstract}
We study the
dynamics of adatoms in a model of vicinal (11$m$) fcc metal surfaces.
We examine the role of different diffusion mechanisms
and their implications to surface growth.
In particular, we study the effect of steps and kinks
on adatom dynamics. We show that the existence of kinks
is crucially important for adatom motion along and across steps.
Our results are in agreement with recent experiments on Cu(100)
and Cu(1,1,19) surfaces. The results also suggest
that for some metals exotic diffusion mechanisms may be important for
mass transport across the steps.

\end{abstract}
\pacs{PACS numbers: 61.50.Cj, 68.35.Fx, 68.55.Bd}

%================================================================ INTRODUCTION

Diffusion of adatoms on solid surfaces is an extensively studied
subject \cite{Diff}. In particular, adatom
dynamics on vicinal metal and semiconductor surfaces has important
implications to surface growth under non--equilibrium conditions
\cite{Grow,Er94}. However, barring a few special cases, the atomistic
details of diffusion processes near steps and kinks are not
known. Current experimental techniques are now able to yield
atomistic information about adatom dynamics near
steps and kinks \cite{Gi93}, and growth of surfaces \cite{Er94,Wu93}.
Clearly,
careful microscopic calculations are needed to understand these phenomena.

Our aim in this Letter is to study models of surfaces of
fcc metals vicinal to the (100) plane.
The open structure of the (100) facets can be expected to
give rise to some unconventional diffusion processes
that are not seen for instance
on stepped surfaces with fcc(111) terraces.
First, we want to identify the various microscopic
mechanisms relevant to self--diffusion near steps and kinks.
Second, we shall discuss the implications of our
results to the morphological stability of
these surfaces under growth \cite{Kr93}, and suggest a
phenomenological model for step growth.
Our results are consistent with experiments
on Cu(100) \cite{Er94} and Cu(1,1,19) \cite{Gi93} surfaces.

%=========================================================== MODEL AND METHODS

The geometric
structure of an fcc(119) surface is shown in Fig.~\ref{fig:view}.
An ideal fcc(11$m$) facet, with odd $m>1$, consists of
(100) terraces of width $ (m-1)r_{nn}/2 $ separated by
(111) steps of height $ r_{nn}/\sqrt{2} $, where
$r_{nn}$ is the distance between nearest neighbor atoms.
Due to the geometry, only one kind of steps (of monolayer height)
with close-packed edges exist on these surfaces.
The metallic interactions between atoms in our model
are derived from the semi--empirical Effective Medium Theory (EMT).
The formalism of EMT is presented in Ref.~\cite{Ja87},
and a description of the implementation
for molecular dynamics (MD) simulations of the present work
can be found in Ref.~\cite{Ha92}.
In the case of copper, EMT has been shown to give a reasonably
accurate quantitative description of many different surface phenomena
\cite{Ja87,Ha92,Ha91,St94,Me94}, which motivates its use for the
present case.

%=============================================== RESULTS - STANDARD MECHANISMS

We shall divide the discussion of the microscopic mechanisms near step
edges into three parts:
standard hopping events (denoted by H),
exchange and other exotic mechanisms (X),
and the effect of kinks on diffusion near and across step edges (K).
For each mechanism M the activation barrier is denoted by $E_{{\rm M}}$,
and that of the reversed process by $E_{{\rm M}}^{\rm rev}$.

In Fig.~\ref{fig:map}(a) we show a contour plot of the
adiabatic energy surface $E(x,y)$ experienced by a single
adatom on the Cu(119) surface at zero temperature.
The potential across the terrace
is shown in Fig.~\ref{fig:map}(b), indicating
the activation energy for diffusion on the terrace $ E_{{\rm A}} $,
and the height of the Schwoebel step barrier
$E^{{\rm rev}}_{{\rm H1}}$ \cite{Sc69}.
It is evident that the barrier
in the $x$ direction is appreciably modulated only at the immediate
vicinity of the steps \cite{Wa93}.
We have verified this for the Cu(1,1,15) surface also.

Activation energies for simple hopping mechanisms
on surfaces of several fcc metals with different orientations,
as given by EMT,
have been extensively tabulated in Ref.~\cite{St94}.
Our results for copper are fully consistent.
The barrier height for a single jump on a flat terrace far from step
edges is found to be $ E_{{\rm A}} = 0.399 $ eV,
and that for diffusion of a vacancy in the first layer of
the terrace is $ E_{{\rm V}} = 0.473 $ eV.
The lowest barrier is that of an adatom diffusing
along a straight step edge, and has a value $ E_{{\rm H2}} = 0.258 $ eV.
As expected, on Cu(11$m$) surfaces we
find $ E_{{\rm H2}} < E_{{\rm A}} < E_{{\rm H1}}^{\rm rev} < E_{{\rm H1}} $.
The inequality $ E_{{\rm H2}} < E_{{\rm A}} $ is consistent
with experimental results \cite{Br93}.

%================================================= RESULTS - EXOTIC MECHANISMS

In addition to ground--state calculations
we have performed extensive MD simulations \cite{MoDy}
to identify possible exotic diffusion mechanisms \cite{Fe90} and
to study entropic contributions to the rates \cite{Me95}.
A well--known mechanism for step crossing,
the replacement of an edge atom by an adatom
from the terrace (mechanism X1)
is observed in our simulations.
In our model the activation energies for
hopping and the simple exchange across the step edge are approximately equal:
$ E_{{\rm H1}} \approx E_{{\rm X1}} $ \cite{Ti93,OneE}.
We have also found more complicated mechanisms for step crossing.
In Fig.~\ref{fig:snap} we show two examples: a ``coherent'' chain
transfer and an atom--by--atom replacement mechanism (vacancy diffusion).
A possible explanation for the first mechanism is
the local close--packed--like order of the surface atoms
in the second configuration of Fig.~\ref{fig:snap}(a),
which may lower the local free energy.
This is obviously an effect characteristic to stepped surfaces with
fcc(100) terraces: e.g.\ on an fcc(111) terrace
the atomic rows are more densely packed and
cannot easily slide with respect to each other.
The second process shown in Fig.~\ref{fig:snap}(b)
can be described as a ``popping up'' of a surface atom
onto the step edge and the diffusion of a {\it vacancy} towards
the descending step. By reaching the step,
the vacancy turns into a hole or a pair of kinks at the step edge,
which can then be filled by a surface atom or an adatom from the terrace
below. Repeating this procedure,
e.g.\ under an external field driving the atoms into the negative $x$
direction, can result in mass transfer across steps which
can then enhance growth instability \cite{Kr93}.

The activation energy of the
first stage of the process shown in Fig.~\ref{fig:snap}(b),
i.e. the pop--up of a surface atom,
was reduced by the existence of a kink at the step edge.
For the processes with and without a kink we find
$ E_{{\rm K3}} < E_{{\rm X3}} $, respectively (cf.\ Fig.~\ref{fig:mech}).
Indeed,
the effect of kinks on the energetics of diffusion processes near step
edges seems to be crucial.
In Fig.~\ref{fig:mech} we show the most important
diffusion routes near straight and kinked step edges.
It turns out
that the hopping of a single adatom {\it across} the step edge in the
vicinity of a kink
site (K1) is not much more favorable than climbing
across a straight edge (H1). On the other hand,
activation barrier for the escape of an adatom from a kink (K4)
is higher than $E_{\rm A}$,
while going ``around the corner'' {\it along} a kinked step edge (K2) is
even more expensive,
i.e.\ $ E_{{\rm H2}} < E_{{\rm A}} < E_{{\rm K4}} < E_{{\rm K2}} $. From
experiments on Cu(1,1,19), the activation energy
for mass transport along kinked step edges was determined
to be $10300 \pm 1630$ K \cite{Gi93}, which was
assumed to be due to the process K4.
For K4 we obtain 6011 K in agreement with Ref. \cite{Ti93}.
However, if we assume that K2 is the rate--limiting
process we get an activation energy of
$E_{{\rm K2}} = 9075$ K which is within the experimental error bars.

%================================================== CONCLUSIONS: STEP CROSSING

The activation barriers for the processes shown in Fig.~\ref{fig:mech}
for Cu(11$m$) are summarized in Table~\ref{table1}.
For any mechanism of step crossing in either direction,
the barrier height is found to be well above $ E_{{\rm A}} $,
i.e.\ a clear Schwoebel barrier exists.
Our results thus indicate that under growth conditions the
currents should go upward which makes the (100) surface
unstable \cite{Kr93}. This is consistent with the
experimental results of Ref.~\cite{Er94} on growth of the Cu(100)
surface.

In the case of our copper model,
the activation barriers for H1, X1, and K3
across the step edge are almost the same.
This means that a finite density of kinks
promotes step crossing.
Due to small differences in activation energy,
the relative occurrence of the different mechanisms at low temperatures
is expected to be strongly model and material dependent.
In particular, our results suggest the possibility
that for (11$m$) surfaces of other fcc metals, exotic
mechanisms such as K3 could play a more important role in mass
transport across step edges.
In addition, some processes such as X2
are influenced by the step density.
More systematic studies on the
effects of inclination and finite temperatures
will be published elsewhere \cite{Me95}.

%================================================== CONCLUSIONS: GROWTH MODELS

As already mentioned,
our results have important implications for growth processes on
copper surfaces vicinal to the (100) plane.
First, under molecular beam epitaxy (MBE)
conditions step crossing should be very rare.
Thus, our data can be used to construct a growth model for
individual steps. In the case of copper,
such a growth model \cite{Me95} should include the following features:
adatom motion along a straight step edge which is very fast,
and motion through a kink site at the step edge which in
turn is much slower than simple diffusion on a flat terrace.
In the simplest approximation,
step crossing and evaporation can be neglected.
The activation energies for these processes are well separated from
each other and well below those of the neglected ones,
which makes the model simple and hopefully applicable to a variety
of vicinal fcc metal surfaces.
This is supported by experiments \cite{Er94,Gi93,Br93}.
For some fcc metals step crossings could be more significant,
and then the detailed
interplay between different mechanisms at step edges and
kink statistics has to be taken into account.
Note that during growth a non--vanishing kink concentration
is naturally provided by adatoms deposited on the terraces.
The growth rules suggested above differ substantially from those expected
to describe MBE on stepped surfaces of silicon \cite{Al95},
where the strong anisotropy of diffusion
\cite{Ro91} results in an interesting step morphology \cite{Wu93,Al95}.

%============================================================ ACKNOWLEDGEMENTS

Discussions with I. Bukharev and H. H\"akkinen
are gratefully acknowledged. J. Krug and S. C. Ying are
acknowledged for a critical reading of the manuscript.
This work has been in part supported by a joint grant
between the Academy of Finland
and Deutcher Akademischer Austauschdienst (DAAD).

%================================================================== REFERENCES

%============================================================= FIGURE CAPTIONS

%===================== FIG.1
\begin{figure}
\caption{Ideal fcc(119) surface: (a) Perspective view and (b) top view.
The size of the unit cell is shown by a dashed line.}
\label{fig:view}
\end{figure}

%===================== FIG.2
\begin{figure}
\caption{The adiabatic
potential experienced by an adatom on the Cu(119) surface as
given by EMT at zero temperature:
(a) a contour plot of the potential, where
the global minimum is at (0,0) and the energy difference between
each contour is 0.1 eV, and
(b) minimum energy route of an adatom diffusing in the
$x$ direction.}
\label{fig:map}
\end{figure}

%===================== FIG.3
\begin{figure}
\caption{Snapshots of two exotic diffusion events from MD
simulations of the Cu(119) surface
(a) at $ T=700 $ K and (b) at $ T=750 $ K. These events only
include surface atoms which are colored black.
Atoms in the adjacent layers are shown in grey.
Only part of the simulation cell is shown. In (a), the surface
atom marked with a cross
is pushed up and a hole is left behind at the descending
step edge. In (b), the surface atom marked with a cross is pushed up and
the vacancy left behind diffuses to the lower step on the right.}
\label{fig:snap}
\end{figure}

%===================== FIG.4
\begin{figure}
\caption{Dominant diffusion mechanisms at the step edge on an fcc(11$m$)
surface (top view).
Black circles are adatoms and open circles denote surface atoms.
A shaded circle shows the position of an atom after the diffusion event
has taken place.
K1 is the mechanism with the lowest activation barrier
of the several possibilities for hopping across a step near a kink.
The five stages of the process X3 are shown by numbers in parentheses.
The processes seen in Fig.~3(a) and 3(b) correspond to X2 and X3,
respectively, albeit without an initial adatom on the lower terrace.}
\label{fig:mech}
\end{figure}

%============================================================== TABLE CAPTIONS

%===================== TABLE I
\begin{table}
\caption{Activation energies of some diffusion mechanisms near step edges
on Cu(11$m$) surfaces as given by EMT. The whole system around the adatom
was allowed to relax in the calculations.
The labels in the first column are those used in Fig.~4.
The third column shows the barriers of the corresponding reversed processes.
For the mechanism X3 the height of the
highest barrier (stage 1) is given.
\label{table1}}
\begin{tabular}{crr}
M (mechanism)&$E_{{\rm M}}$ (eV)&$E_{{\rm M}}^{\rm rev}$ (eV)\\
\tableline
H1&0.867&0.578\\
H2&0.258&$0.258$\\
\tableline
X1&0.909&0.631\\
X2&1.310&$1.310$\\
X3&1.074&$-$\\
\tableline
K1&0.842&0.573\\
K2&0.782&0.492\\
K3&0.879&$0.143$\\
K4&0.518&0.239\\
\end{tabular}
\end{table}

%===================================================================== THE END


\begin{references}
\bibitem{MoDy}
For Cu(119) surface we have simulation
cells of the sizes $ (N_{s},L_{y},L_z) = (6,10,12) $
and  $ (N_{s},L_{y},L_z) = (4,20,12) $,
where $N_{s}$ is the number of steps in the cell,
$L_{y}r_{nn}$ is the size of the cell in $y$ direction, and $L_z$ is
the number of atomic layers.
In simulations,
the feasible temperature range was between 700--800 K, and
the simulation length
was of the order of a few nanoseconds.
Computational details will be reported elsewhere \cite{Me95}.
\end{references}
\end{document}